# THE DOPING PHASE DIAGRAM OF $Y_{1-x}Ca_xBa_2(Cu_{1-y}Zn_y)_3O_{7-\delta}$ FROM TRANSPORT MEASUREMENTS: TRACKING THE PSEUDOGAP BELOW $T_c$ (y = 0)

S.H. Naqib[1]*, J.R. Cooper[1] and J.L. Tallon[2]

[1]IRC in Superconductivity, University of Cambridge, Madingley Road, Cambridge CB3 OHE, UK
[2]MacDiarmid Institute for Advanced Materials and Nanotechnology, Victoria University and Industrial Research Ltd., P.O. Box 31310, Lower Hutt, New Zealand
*Corresponding author: Phone: +44–1223-337049; Fax: +44-1223-337074; E-mail: shn21@hermes.cam.ac.uk

## Abstract

The effects of planar hole concentration, p, on the resistivity, ρ(T), of sintered $Y_{1-x}Ca_xBa_2(Cu_{1-y}Zn_y)_3O_{7-\delta}$ samples were investigated over a wide range of Ca, Zn, and oxygen contents. Zn was used to suppress superconductivity and this enabled us to extract the characteristic pseudogap temperature, $T^*(p)$, from ρ(T,p) data below $T_{co}(p)$ [ ≡ $T_c$ (y = 0)]. We have also located the characteristic temperature, $T_{scf}$, marking the onset of significant superconducting fluctuations above $T_c$, from the analysis of ρ(T,H,p) and ρ(T,p) data. This enabled us to identify $T^*(p)$ near the optimum doping level where the values of $T^*(p)$ and $T_{scf}(p)$ are very close and hard to distinguish. We again found that $T^*(p)$ depends only on the hole concentration p, and not on the level of disorder associated with Zn or Ca substitutions. We conclude that (i) $T^*(p)$ (and therefore, the pseudogap) persists below $T_{co}(p)$ on the overdoped side and does not merge with the $T_{co}(p)$ line and (ii) $T^*(p)$, and thus the pseudogap energy, extrapolates to zero at the doping p = 0.19 ± 0.01.

# 1. INTRODUCTION

One of the most remarkable phenomena in high-$T_c$ cuprates is the observation of the pseudogap in the spectra of charge and spin excitations [1-3]. The pseudogap (PG) is detected by various experimental techniques [1-3] over a certain range of planar hole concentrations, p (the number of added holes per $CuO_2$ plane), extending from the underdoped (UD) to the slightly overdoped (OD) regions. In the 'pseudogap state' various anomalies are observed both in the normal and superconducting (SC) states, which can be interpreted in terms of a drop in the single particle density of states [3]. A number of theoretical explanations have been proposed for the origin for the PG, which is believed to be an essential feature of the physics of the normal state (NS) as well as the SC state of the cuprates. Existing theories of the PG can be classified broadly into two categories. The first is based upon incoherent Cooper pair formation for $T < T^*$ well above the SC transition temperature, $T_c$, [4-6] with long-range phase coherence appearing only at $T \leq T_c$. In this scenario $T^*$ may be the mean-field $T_c$. $T^*(p)$ merges with the $T_c(p)$ phase curve in the OD region ($p \geq 0.21$) where the pair formation temperature is essentially the same as the phase coherence temperature. In the second scenario, the PG is ascribed to fluctuations of some other type that coexist and generally compete with superconductivity. The most popular picture here is that of antiferromagnetic (AFM) fluctuations, but similar effects have been attributed to charge density waves, a structural phase transition or electronic phase separation on a microscopic scale (e.g., the stripe scenario) [7-10]. One key constraint, which may rule out many of the above simple models, is that the single-particle density of states has a mysterious 'states non-conserving' property in the PG state [11]. The experimental situation is often thought to be rather inconclusive regarding the origin of the PG [1,2].

The evolution of ρ(T) with p provides a way of establishing the T-p phase diagram of high-$T_c$ superconductors (HTS) and can give estimates of $T^*(p)$, which is identified from a characteristic downturn in ρ(T). One particular problem associated this method is that when $p \approx p_{opt}$ (where $T_c(p)$ is maximum), $T^*$ is close to the temperature, $T_{scf}$, at which the effect of SC fluctuation is clearly seen in ρ(T). This is not a problem for theories belonging to the first group where $T^*(p)$ itself is essentially derived from strong

SC fluctuations. For the second scenario, this poses a problem as SC fluctuations (and superconductivity itself) mask the signatures of the predicted PG in the vicinity of (and below) $T_c$. With the notable exception of specific heat measurements [3], most experimental techniques are unable to track the predicted $T^*(p)$ below $T_c(p)$. One way of avoiding this problem is to suppress superconductivity with a magnetic field and reveal the 'NS' below $T_c$, because the PG is comparatively insensitive to magnetic field [12]. In practice this is very difficult because of the very large upper critical field, $H_{c2}$, of many HTS. The other way is to destroy superconductivity by introducing disorder, e.g., by alloying with Zn. Because of the d-wave order parameter, Zn substitution suppresses superconductivity most effectively and, like a magnetic field, has little effect on $T^*(p)$ [13]. We took this second route to look for $T^*(p)$ below $T_{co}(p)$.

In this paper we report a systematic study of the transport properties of the superconducting compound $Y_{1-x}Ca_xBa_2(Cu_{1-y}Zn_y)_3O_{7-\delta}$. We have measured resistivity, room-temperature thermopower, S[290K], and the AC susceptibility (ACS) of a series of sintered $Y_{1-x}Ca_xBa_2(Cu_{1-y}Zn_y)_3O_{7-\delta}$ samples with different levels of Zn, Ca, and oxygen contents. The motivation for the present study is to locate $T^*(p)$ from the evolution of $\rho(T)$ of Y123 with different amounts of Zn and Ca extending from UD to OD states. While pure Y123 with full oxygenation ($\delta = 0$), is slightly OD, further overdoping can be achieved by substituting $Y^{3+}$ by $Ca^{2+}$. The advantages of using Zn are: (i) it mainly substitutes the in-plane Cu(2) sites, thus the effects of planar impurity can be studied and (ii) the doping level remains nearly the same when Cu(2) is substituted by Zn, enabling one to look at the effects of disorder on various normal and SC state properties at almost the same hole concentration [14,15]. From the analysis of $\rho(T,p)$ and $\rho(T,p,H)$ data we extracted the p-dependence of the characteristic temperatures $T^*$ and $T_{scf}$, and find that indeed T* falls below $T_{co}$ providing strong evidence for the second scenario, namely that the PG arises from an independent coexisting correlation.

## 2. EXPERIMENTAL DETAILS AND RESULTS

Single-phase polycrystalline sintered samples of $Y_{1-x}Ca_xBa_2(Cu_{1-y}Zn_y)_3O_{7-\delta}$ were synthesized in both laboratories by solid-state reaction methods using high-purity (>

99.99%) powders. The details of sample preparation and characterization can be found in refs. [16,17]. The NS and SC properties, including T*, of HTS are highly sensitive to p and, therefore, it is important to determine p as accurately as possible. The room temperature thermopower, S[290K], has a systematic variation with p for various HTS over the entire doping range extending from very UD to heavily OD regimes [18], also S[290K] is insensitive to in-plane disorder like Zn and the crystalline state of the sample [19]. For these reasons S[290K] is a good measure of p even in the presence of strong in-plane scattering by $Zn^{2+}$ ions. For all our samples we have used S[290K]$^{\otimes}$ [20] to determine p. Using these values of p, we find that the parabolic $T_c(p)$ formula [21], given by

$$\frac{T_c(p)}{T_c(p_{opt})} = 1 - Z(p - p_{opt})^2 \qquad (1)$$

is obeyed for all samples. Usually, for Zn-free samples, Z and $p_{opt}$ seem to take on universal values but these parameters increase systematically with increasing Zn content [22]. In the present case, Z increases from the usual 82.6 [21] for the Zn-free sample to 410 for 6%-Zn sample [22] and $p_{opt}$ rises from 0.16 for the Zn-free compound [21] to 0.174 for the 6%Zn-20%Ca sample [22]. The physical meaning of these changes in the fitting parameters is that as the Zn content is increased, the region of superconductivity shrinks and becomes centred on higher values of p, before finally forming a small bubble around p ~ 0.19 and disappearing completely for y ~ 0.1 [23]. From the observed evolution of $p_{opt}(y)$ we previously inferred that superconductivity for this system is at its strongest at p ~ 0.185 (Fig.1 of ref.[22]), as this remains the last point of superconductivity at a critical Zn concentration (defined as the highest possible Zn concentration for which superconductivity just survives, considering all p-values). This point has been made earlier in other studies [22,23] and the value p ~ 0.19 is indeed a special value at which the PG vanishes quite abruptly, as seen from the analysis of specific heat data of HTS cuprates [11,23].

---

$^{\otimes}$S[290K] = -139p + 24.2           for p > 0.155

  S[290K] = 992exp(-38.1p)       for 0.155 ≥ p > 0.05

The hole concentration was varied by changing both the oxygen deficiency and the Ca content. We have obtained $T_c$ from both resistivity and low field ($H_{rms}$ = 0.1Oe; f = 333.3Hz) ACS measurements. $T_c$ was taken at zero resistance (within the noise level of ± $10^{-6}$ Ω) and at the point where the line drawn on the steepest part of diamagnetic ACS curve meets the T-independent base line associated with the negligibly small NS signal. $T_c$-values obtained from these two methods agree within 1K for most of the samples [22]. We placed particular emphasis on the determination of p and $T_c$ as accurately as possible because of the extreme sensitivity of various ground-state SC and NS properties to p. This is especially true near the optimum doping level, where, although $T_c$(p) is nearly flat, the SC condensation energy, superfluid density, PG energy scale etc. change quite abruptly and substantially for a small change in p [11,23-25].

Sintered bars of 89 to 93% of the theoretical density were used for resistivity measurements. Resistivity was measured using the four-terminal method with an ac current of 1 mA (77 Hz) using 40 μm dia. copper wire and silver paint to make the contacts. Typical contact resistances were below 2Ω. Transformers were used to minimize electrical noise. We have tried to locate the PG temperature, $T^*$, with a high degree of accuracy. At this point we should emphasise that $T^*$(p) does not represent a phase transition temperature but instead $k_B T^*$(p) is some kind of a characteristic energy scale of a lightly-doped Mott-insulator, probably reflecting the energy of correlated holes and spins [11]. Plots of dρ(T)/dT vs. T and ρ(T) - $ρ_{LF}$ vs. T yield very similar $T^*$ values (within ± 5K) ($ρ_{LF}$ is a linear fit $ρ_{LF}$ = b+cT, in the high-T linear region of ρ(T), above $T^*$) [22]. Using $T^*$/T as a scaling parameter, it is possible to normalize ρ(T). The result of the scaling was shown in ref. [22], where, leaving aside the SC transitions, all resistivity curves collapsed reasonably well on to one p-independent universal curve over a wide temperature range. It is important to note that Zn does not change the $T^*$(p) values but it suppresses $T_c$(p) very effectively. For example, $T^*$(p ~ 0.115) ~ 250 ± 5K for both Zn-free and the 3%-Zn samples but $T_c$ itself is suppressed from 70K (Zn-free) to 29K (3%-Zn) [22]. Similar results have been obtained for Zn-substituted Y123 by other studies [13]. This very different p-dependence for $T_c$ and $T^*$ has often been stated as an indication for the non-SC origin of the PG [26].

Magneto-transport measurements were made using an *Oxford Instruments* superconducting magnet system. The sample temperature was measured using a *Cernox* thermometer with an absolute accuracy of ~ 50mK. We have located $T_{scf}$, from the analysis of $\rho(T,H)$ and $d^2\rho(T)/dT^2$ vs. T data, in the T-range from $T_c$ to ~ $T_c$ + 30K. Here we have used the facts that (i) $\rho(T,H)$ becomes strongly field sensitive below $T_{scf}$, where experimentally we find that conventional SC fluctuations set in quite abruptly, but $T^*$ is unaffected by magnetic field, at least for H < 10 Tesla [12] and (ii) $\rho(T)$ shows a much stronger, and progressively accelerating, downturn at $T_{scf}$ than that present at $T^*$. Plots of $d^2\rho(T)/dT^2$ vs. T mask the PG feature and visually enhance the effects of SC fluctuations near $T_{scf}$. This is clearly seen in Fig.1, where $\rho(T)$ of a Zn-free UD sample (p ~ 0.124 and $T^*$ ~ 230K) has been well fitted to the form $a+bT+cT^2$ in the temperature range of 300K > T > $T_{scf}$. Figs.2 show the results of analysis of $\rho(T)$ data for the extraction of $T_{scf}$ for some of the samples. Three different methods (including subtraction of a linear fit of $\rho(T)$ to locate the onset of the strong deviation in $\rho(T)$ from linearity near the superconducting transition) are applied and all of them give the same value of $T_{scf}$ to within ± 2.5K.

Fig.3 shows $T_c(p)$ and $T_{scf}(p)$ of two representative 20%-Ca samples with 0%-Zn and 4%-Zn. $\Delta T_{scf}(p)$ is relatively insensitive of Zn content (y) (see the inset of Fig.3), this quantity was also found to be independent of Ca content (x) and crystalline state of the samples (analysis of published $\rho(T)$ data of single crystal c-axis films of $Y_{1-x}Ca_xBa_2Cu_3O_{7-\delta}$ [27] and $YBa_2(Cu_{1-y}Zn_y)_3O_{7-\delta}$ [28] gave values of $\Delta T_{scf}(p)$ to within ± 2K, of those found for our sintered polycrystalline samples). We have tried to model the evolution of $\Delta T_{scf}(p)$ roughly from the temperature dependence of $\rho$ at high temperatures and from the form of the Aslamazov - Larkin (AL) contribution to the fluctuation conductivity [29]. We show this in Fig.4. The excess conductivity due to superconducting fluctuation, $\Delta\sigma(T)$, can be defined as, $\Delta\sigma(T) = 1/\rho(T) - 1/\rho_{BG}(T)$, where, $\rho_{BG}(T)$ is the background resistivity taken as the high-T fit (T > $T_c$ + 50K) of the $\rho(T)$ data. The AL contribution to the fluctuation conductivity for a two-dimensional superconductor, $\Delta\sigma^{AL}$, is expressed as, $\Delta\sigma^{AL} = [e^2/(16\hbar d)]\varepsilon^{-1}$ [29], where $\varepsilon = \ln(T/T_c)$ and d is the periodicity of superconducting layers (d ~ 11.7Å for Y123, assuming that the $CuO_2$ bilayers fluctuate as one unit). In this analysis we have taken $T_c$ at zero resistivity (within the noise level).

Fig.4 shows the T-dependence of $d^2\rho(T)/dT^2$ and $d^2\rho_{tot}(T)/dT^2$, with $1/\rho_{tot}(T) = 1/\rho_{BG}(T) + \Delta\sigma^{AL}(T)$. In the absence of a proper energy or a momentum cut-off [30,31], extrinsic effects due to the grain boundary resistance in polycrystalline samples (which is included in $\rho_{BG}$), and problems associated with proper identification of the mean field SC transition temperature for samples with finite transition width, a good quantitative agreement between $\rho_{tot}(T)$ and $\rho(T)$ cannot be expected [30], but nevertheless $d^2\rho(T)/dT^2$ and $d^2\rho_{tot}(T)/dT^2$ show remarkable agreement (see Fig.4). We have made some model calculations suggesting that as long as the grain boundary resistance remains temperature independent, it does not effect the second derivative significantly. In an effort to understand the systematic trends in $\Delta T_{scf}$ shown in the inset to Fig. 3 we have examined various terms in the second temperature derivative of $\rho_{tot}(T)$ as follows, writing

$$\frac{1}{\rho_{tot}(T)} = \frac{1}{\rho_{BG}(T)} + C\varepsilon^{-1} \qquad (2)$$

where $C = e^2/(16\hbar d)$, we obtain (ignoring the insignificant terms)

$$\frac{d^2\rho_{tot}}{dT^2} \approx C\left(\frac{\rho_{tot}}{\varepsilon}\right)^2 \left(\frac{1}{T^2}\right)\left\{-\frac{2}{\varepsilon} - 1\right\}$$

now, assuming $-\eta$ be the threshold value of the second derivative for which we can identify a characteristic temperature $T = T_{scf}$ then the above equation becomes, after some rearrangements

$$\frac{\varepsilon_{scf}^3}{2 + \varepsilon_{scf}} T_{scf}^2 \approx \frac{C\rho_{scf}^2}{\eta} \qquad (3)$$

where, $\varepsilon_{scf} = \ln(T_{scf}/T_c)$ and $\rho_{scf} = \rho_{tot}(T_{scf})$. We have plotted the left-hand side of equation (3) versus experimental values of $\rho^2(T_{scf})$ in Fig.5 using the approximation $\rho_{tot}(T_{scf}) \sim \rho(T_{scf})$. A remarkably linear trend is found and this gives further credence to our fluctuation analysis. The decreasing trend in $\Delta T_{scf}$ with increasing doping shown in Fig. 3 is therefore primarily associated with the decreasing absolute resistivity as summarised by eq. (3).

Once we have located $T_{scf}$, we are in a position to look at $T^*(p)$ (below $T_{co}(p)$) for Zn substituted samples. There is one disadvantage of Zn substitution that can hamper the identification of $T^*$ from $\rho(T)$ measurements, namely, Zn tends to localize low-energy quasiparticles (QP) and induces an upturn in $\rho(T)$. This upturn starts at increasingly higher temperatures as p decreases and, to a lesser extent, as the Zn content increases, and thus can mask the downturn due to the PG at $T^*$ [13,22]. Indeed, the upturn temperature, $T_{min}$, has been found to scale with $T^*$ and is evidently associated with the pseudogap, trending to zero as $p \rightarrow 0.19$ [32]. In this study we have taken care of this fact by confining our attention to the $\rho(T,p)$ data for lower Zn contents in the underdoped region ($p < 0.14$). Fig.6a shows the $\rho(T)$ data of $Y_{0.80}Ca_{0.20}Ba_2(Cu_{0.96}Zn_{0.04})_3O_{7-\delta}$ for $p = 0.175 \pm 0.003$. The insets of Fig.6 clearly show the downturn associated with the PG at around 80K. It should be noted that $T_c$ of this compound is 43K and $T_{scf} = 62.5K$ (see Fig.3). Fig.6b shows a similar analysis for $Y_{0.80}Ca_{0.20}Ba_2(Cu_{0.985}Zn_{0.015})_3O_{7-\delta}$ with $p = 0.178 \pm 0.003$. The $\rho(T)$ features of this compound once again show a $T^* \sim 80K$ with now a $T_c$ of 54K and $T_{scf} = 73K$. Considering the facts that $T^*(p)$ values are the same irrespective of Ca and Zn contents (at least for the level of substitutions used in present study), and $T_{co-max}$ for pure Y123 is 93K, these $T^*(p)$ are substantially below the respective $T_{co}(p)$ values (~ 90.5K) (see Fig.7). This study, to our knowledge, is the first instance where $T^*(p)$ has been tracked down below $T_{co}(p)$ from any transport measurement, although of course this has effectively been done earlier by analysis of specific heat data by Loram et al. [3,11] and NMR data by Tallon *et al*. [33]. In Fig.7 we construct a doping 'phase diagram' for $Y_{1-x}Ca_xBa_2(Cu_{1-y}Zn_y)_3O_{7-\delta}$, including $T^*(p)$ and the PG energy scale from some other sources [23,34]. We have taken $E_g(p)/k_B$ from specific heat measurements [23]. $E_g$ is the energy scale for the PG and $E_g(p)/k_B \sim \theta T^*(p)$, where ($\theta = 1.4 \pm 0.1$) is a certain proportionality constant [26]. The extrapolated $T^*(p)$ goes to zero at $p = 0.19 \pm 0.01$, following the same trend as $E_g(p)/k_B$.

# 3. DISCUSSION

From a careful analysis of the resistivity data we have been able to track $T^*(p)$ below $T_{co}(p)$. At this point we would like to stress that $T^*(p)$ values obtained from $\rho(T)$ are independent of the crystalline state and are the same for polycrystalline and single crystal samples [22]. Considering the disorder (Zn and Ca) independence of $T^*(p)$, our findings confirm that the PG exists in the SC region below $T_c$ of the Zn-free samples and $T^*(p)$ does not merge with $T_c(p)$ in the OD region as proposed by various precursor pairing scenarios for PG. As explained in the present work, if one is unable to distinguish between $T^*(p)$ and $T_{scf}(p)$, one can easily be lead to the wrong conclusion that $T^*(p)$ exists for $p > 0.19$ and merges with $T_c(p)$ on further overdoping. As $T_{scf}$ is ~ $T_c + 22K$ near optimum doping level, the close proximity between $T_{scf}$ and $T^*$ makes an accurate identification of $T^*$ very difficult unless both $T_c$ and thus $T_{scf}$ are suppressed by some means that does not affect $T^*$. Zn serves this purpose very well. Unfortunately, we have been unable to track $T^*(p)$ to below ~ 80K for samples with $p > 0.180$ with enough accuracy. This is because $T^*(p)$ decreases very sharply with increasing p (compared with the decrease in $T_c(p)$ and $T_{scf}(p)$, see Figs.3 and 7) and becomes very close to or even goes below $T_{scf}$ and eventually $T_c(p)$ [3,11]. For example, $Y_{0.90}Ca_{0.10}Ba_2(Cu_{0.985}Zn_{0.015})_3O_{7-\delta}$ with $p = 0.180 \pm 0.0035$ has a $T_c = 64K$ and $\rho(T)$ is linear over 310K to 80K, marked and accelerating downturn in $\rho(T)$ starts at ~ 79K. Considering that, for this hole concentration, we have $\Delta T_{scf} = 17 \pm 2.5$ K for all our samples, independent of Zn concentration, then this downturn at 79K is necessarily that associated with SC fluctuations and $T^*$ must lie below this temperature. Another example is $Y_{0.80}Ca_{0.20}Ba_2(Cu_{0.96}Zn_{0.04})_3O_{7-\delta}$ with $p = 0.184 \pm 0.0035$, which has $T_c = 42.25K$ and $\rho(T)$ is linear over 310K to 60K. A marked and accelerating downturn in $\rho(T)$ starts at ~ 58.5K (= $T_{scf}$). Thus for this compound $T^* < 58.5K$. Although Zn substitutions greater than 4% can help us track $T^*(p)$ below 80K precisely, this will also increase the QP localization at lower temperatures and may mask the PG-feature in $\rho(T)$.

$T_{scf}(p)$ has been taken as the onset temperature of detectable SC fluctuations in the present study. The disorder independence of $\Delta T_{scf}(p)$ (inset of Fig.3) and a simple AL-type analysis of $\rho(T)$ data (see Fig.4) strongly support this assumption. Disorder suppresses both $T_c(p)$ and $T_{scf}(p)$ in exactly the same way, unlike $T^*(p)$ which remains unaffected.

## 4. CONCLUSIONS

In summary, we have analysed our $\rho(T)$ data to determine the $T^*(p)$ of $Y_{1-x}Ca_xBa_2(Cu_{1-y}Zn_y)_3O_{7-\delta}$ over a wide doping range and compositions. We have shown that $T^*(p)$ falls below $T_{co}(p)$ in the OD side, does not merge with $T_c(p)$, and the extrapolated $T^*(p)$ becomes zero at $p = 0.19 \pm 0.01$. We have also extracted a characteristic temperature, $T_{scf}$, at which SC fluctuations become detectable. It is perhaps surprising that $T_{scf}$ is so well-defined experimentally in all our samples, but the very different p and y (Zn) dependence of $T^*$ and $T_{scf}$ points towards their different origins, e.g., $T_{scf}$ is related to precursor SC, whereas $T^*$ has a non-SC origin. Our findings support the picture proposed by Loram et al. based on their specific heat study that the PG vanishes at a critical doping, $p_{crit} \sim 0.19$, and coexists with SC for $p < 0.19$ [11,23]. Recently Krasnov et al. have reached to similar conclusions based on their intrinsic tunneling spectroscopy studies of Bi-2212 single crystals [12,35].

**ACKNOWLEDGEMENTS:** We gratefully acknowledge J.W. Loram for helpful comments and suggestions. One of us (SHN) acknowledges the financial support from the Commonwealth Scholarship Commission (UK), Darwin College, Cambridge Philosophical Society, Lundgren Fund, and the Department of Physics, Cambridge University.

**Figure Captions (S.H. Naqib et al.)**

**[PAPER TITLE: The doping phase diagram of .........]**

Fig.1. $\rho(T)$ data (dashed line) for underdoped $Y_{0.80}Ca_{0.20}Ba_2Cu_3O_{7-\delta}$ ($T_c$ = 77.5K, p = 0.124). The full line represents a fit to $\rho(T) = a+bT+cT^2$.

Fig.2. Determination of $T_{scf}$. Main panel: dashed lines - experimental $\rho(T)$, thick solid lines - $d^2\rho(T)/dT^2$ (LH scale), thin solid lines - linear fits to $\rho(T)$ near the superconducting transition (see text). Insets: magnetoresistance data plotted as $\rho(H = 6$ Tesla$)/ \rho(H = 0$ Tesla$)$ vs. T. (a) $Y_{0.80}Ca_{0.20}Ba_2(Cu_{0.97}Zn_{0.03})_3O_{7-\delta}$, p = 0.185, and $T_c$ = 45.8K and (b) $Y_{0.80}Ca_{0.20}Ba_2(Cu_{0.96}Zn_{0.04})_3O_{7-\delta}$, p = 0.198, and $T_c$ = 36.7K.

Fig.3. Main panel: $T_c(p)$ and $T_{scf}(p)$ for $Y_{0.80}Ca_{0.20}Ba_2(Cu_{1-y}Zn_y)_3O_{7-\delta}$. Inset: $\Delta T_{scf}(p)$ for $Y_{0.80}Ca_{0.20}Ba_2(Cu_{1-y}Zn_y)_3O_{7-\delta}$. Zn-contents (y in %) are shown in the figure.

Fig.4. $d^2\rho/dT^2$ and $d^2\rho_{tot}/dT^2$ vs. T for $Y_{0.90}Ca_{0.10}Ba_2Cu_3O_{7-\delta}$. $\rho$ is the experimental resistivity and $1/\rho_{tot}$ is the sum of the background conductivity and the AL fluctuation conductivity (see text). (a) An overdoped sample ($T_c$ = 78K, p = 0.18) and (b) an underdoped sample ($T_c$ = 83K, p = 0.149). Arrows indicate $T_{scf}$. Insets show $\rho(T)$ near the superconducting transition.

Fig.5. $[(\varepsilon_{scf}^3 T_{scf}^2)/(2 + \varepsilon_{scf})]$ vs. $\rho^2(T_{scf})$ (see eq. (3) in text) for the $Y_{0.80}Ca_{0.20}Ba_2(Cu_{1-y}Zn_y)_3O_{7-\delta}$ samples. Zn-contents (y in %) are shown in the figure. The dashed straight line is drawn as a guide to the eye.

Fig.6. $T^*$ for (a) $Y_{0.80}Ca_{0.20}Ba_2(Cu_{0.96}Zn_{0.04})_3O_{7-\delta}$ (p = 0.175 ± 0.003) and (b) $Y_{0.80}Ca_{0.20}Ba_2(Cu_{0.985}Zn_{0.015})_3O_{7-\delta}$ (p = 0.178 ± 0.003). $T_{scf}$ is also shown in the figure.

Fig.7. Characteristic temperatures, ($T^*$ and $E_g/k_B$) for $Y_{1-x}Ca_xBa_2(Cu_{1-y}Zn_y)_3O_{7-\delta}$: (■) Y123 c-axis thin film [34], (▼)$Y_{0.95}Ca_{0.05}Ba_2Cu_3O_{7-\delta}$, (●)$Y_{0.90}Ca_{0.10}Ba_2Cu_3O_{7-\delta}$, (◆)$Y_{0.80}Ca_{0.20}Ba_2Cu_3O_{7-\delta}$, (▲)$Y_{0.80}Ca_{0.20}Ba_2(Cu_{0.985}Zn_{0.015})_3O_{7-\delta}$, (□)$Y_{0.80}Ca_{0.20}Ba_2(Cu_{0.97}Zn_{0.03})_3O_{7-\delta}$, (O) $Y_{0.80}Ca_{0.20}Ba_2(Cu_{0.96}Zn_{0.04})_3O_{7-\delta}$, (Δ)$Y_{0.80}Ca_{0.20}Ba_2Cu_3O_{7-\delta}$ (specific heat) [23]. The thick dashed line shows $T_{co}(p)$ for pure Y123, drawn using equation (1) with $T_{co\text{-}max} = 93K$. The thick straight line is a guide to the eye.

Fig.1. (S.H. Naqib et al.) [The doping phase diagram of…………]

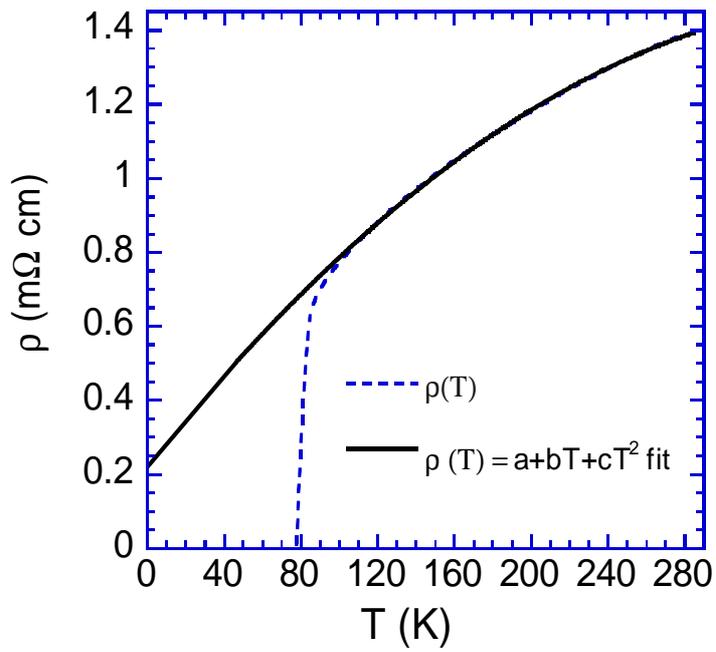

Fig.2. (S.H. Naqib et al.) [The doping phase diagram of……...]

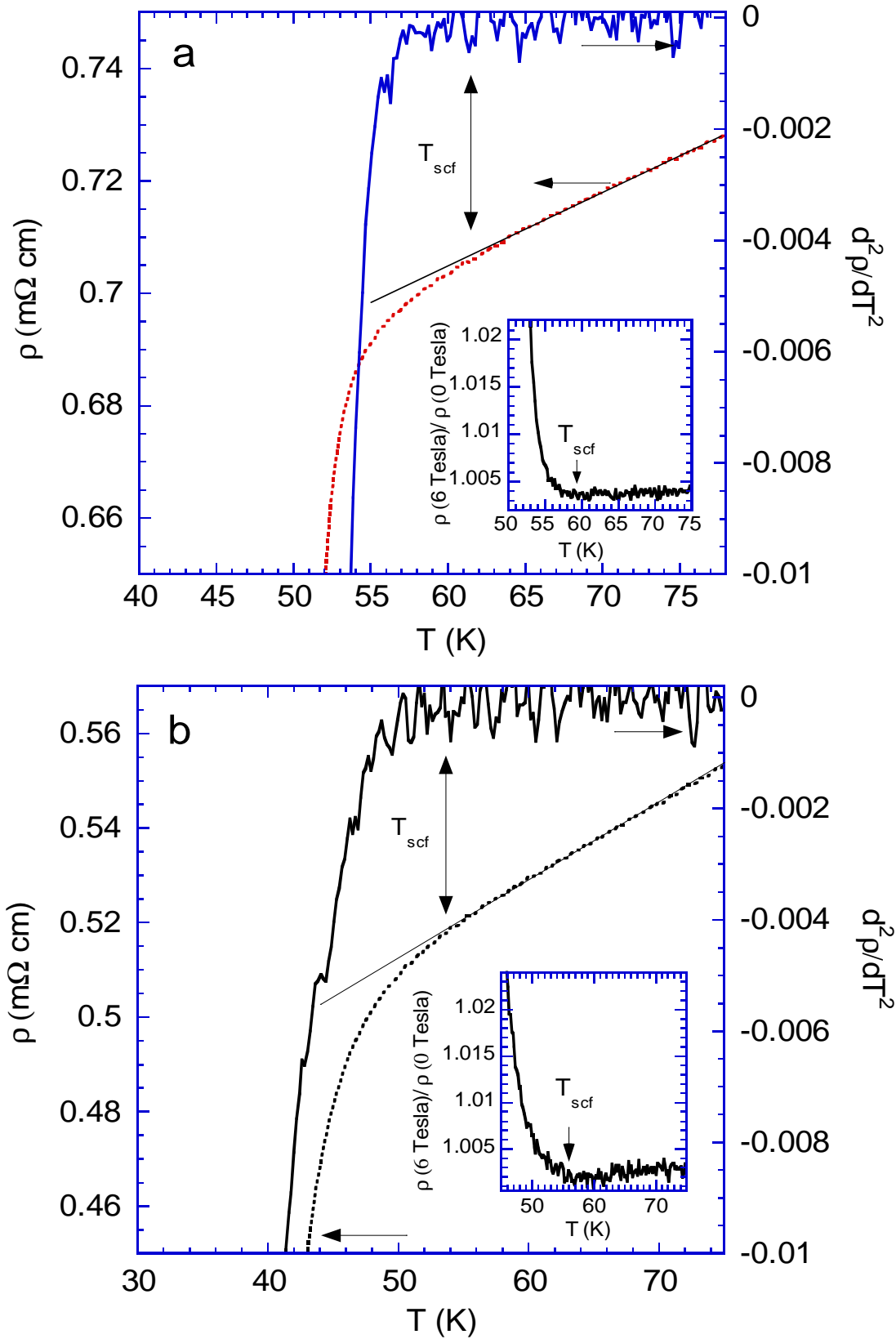

Fig.3. (S.H. Naqib et al.) [The doping phase diagram of………..]

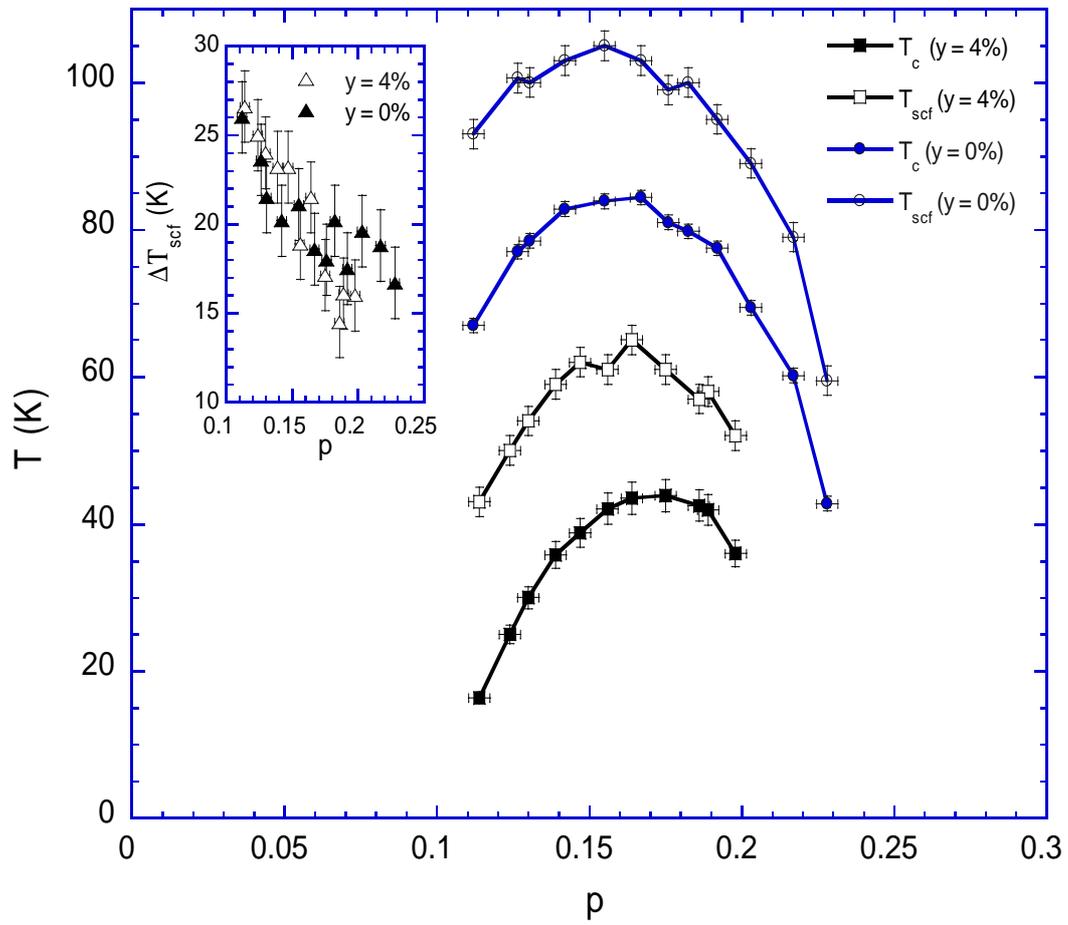

Fig.4. (S.H. Naqib et al.) [The doping phase diagram of.................]

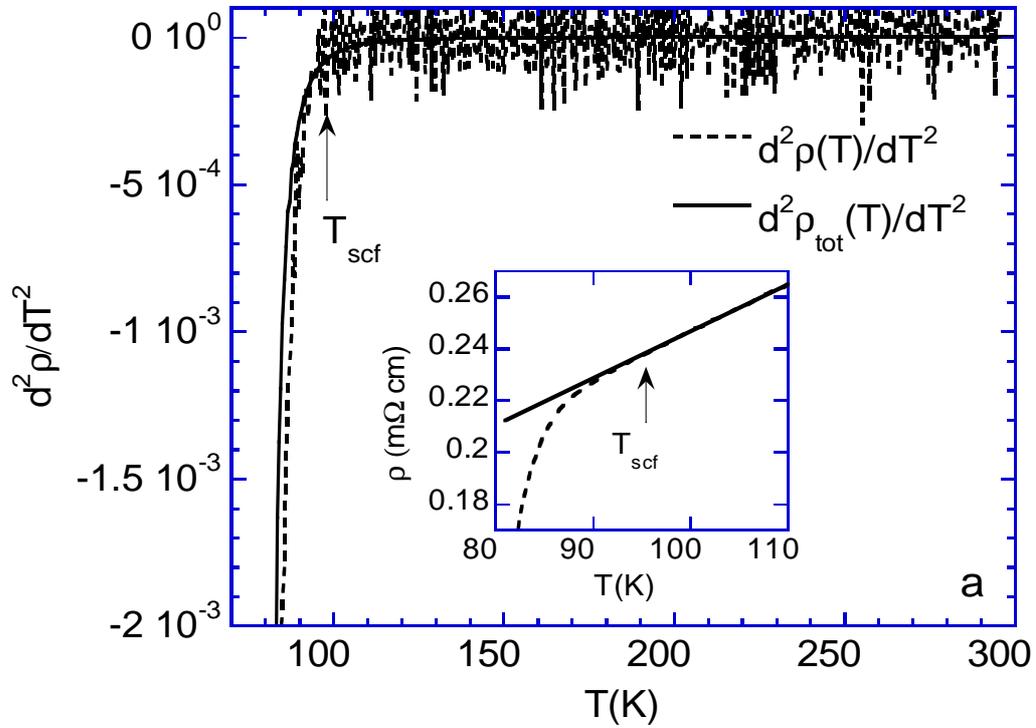

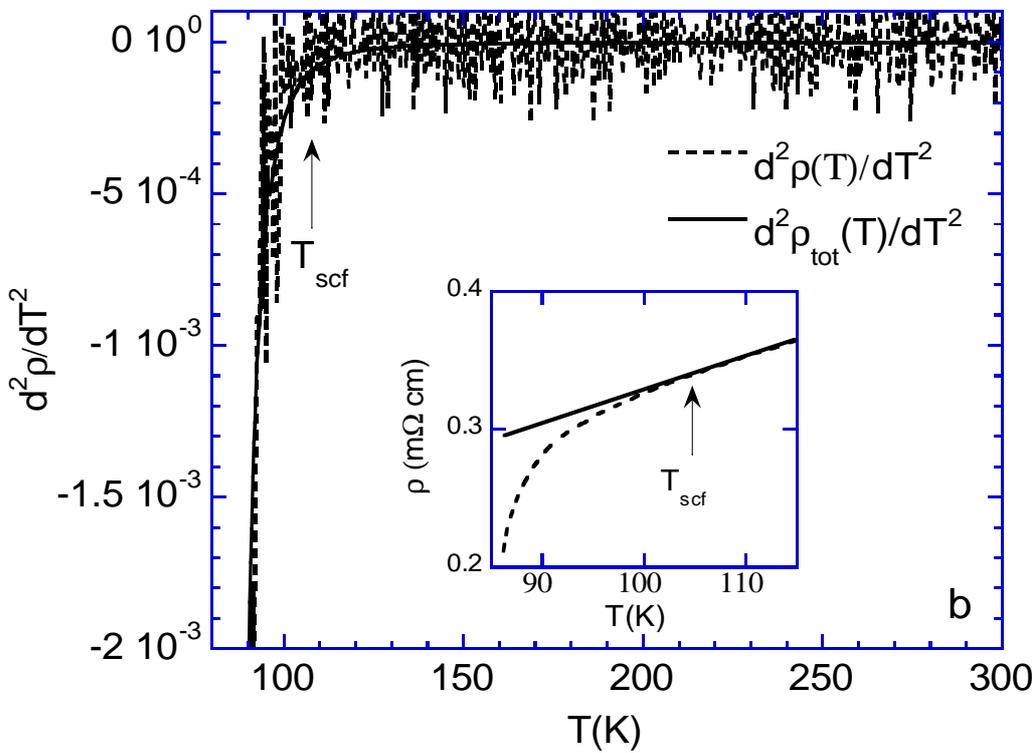

Fig.5. (S.H. Naqib et al.) [The doping phase diagram of............]

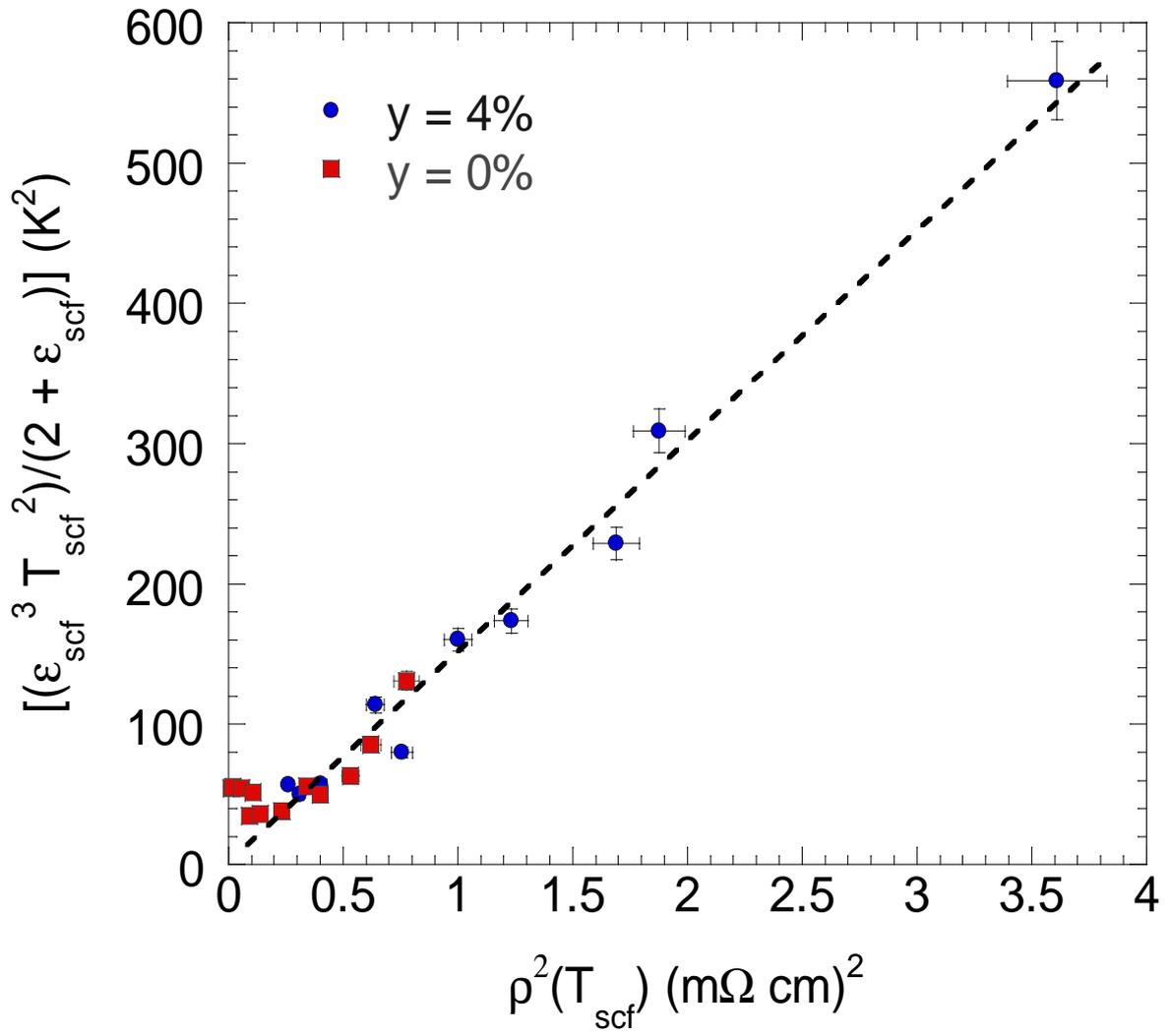

Fig.6. (S.H. Naqib et al.) (The doping phase diagram of……)

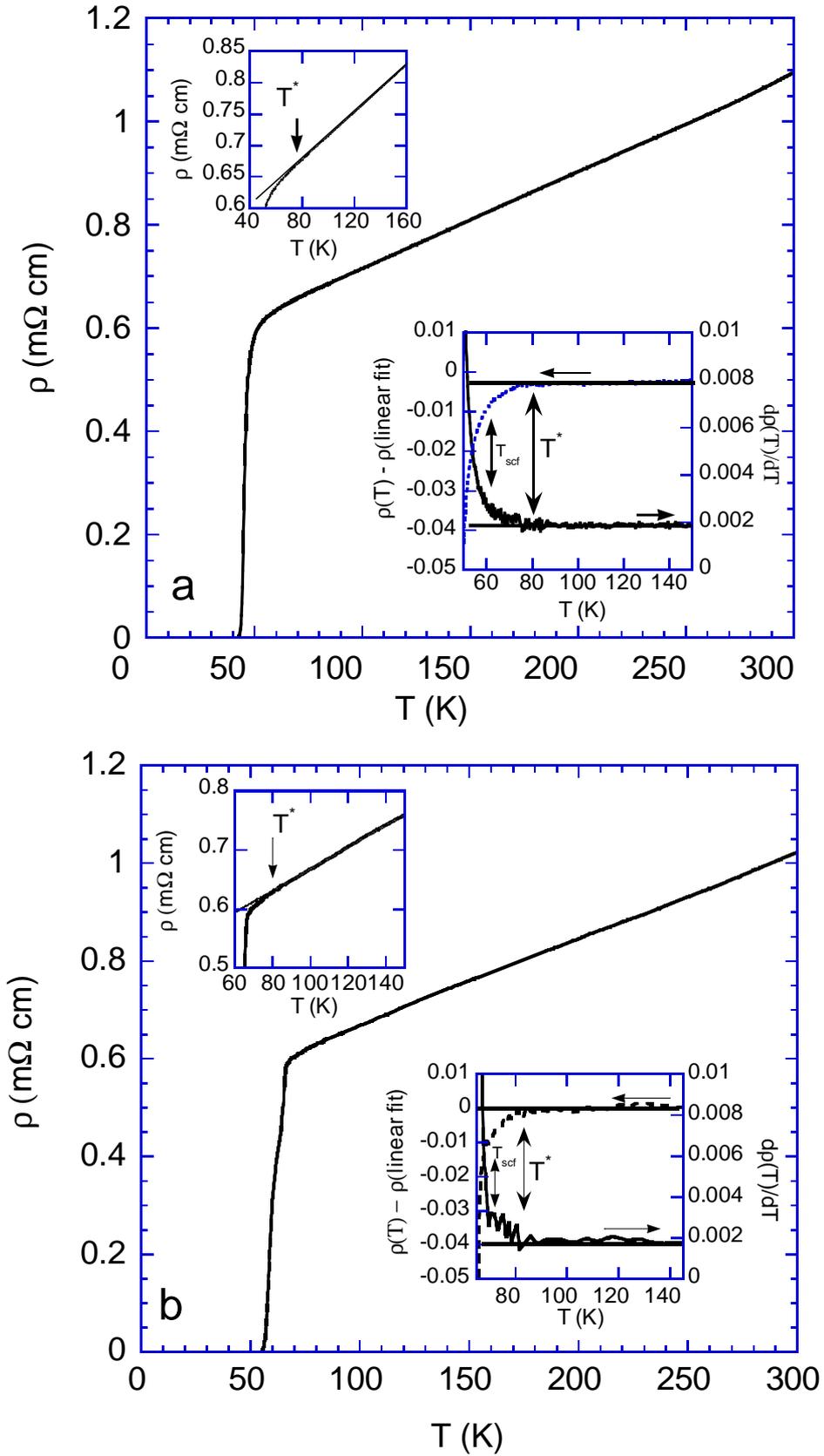

Fig.7. (S.H. Naqib et al.) (The doping phase diagram of……)

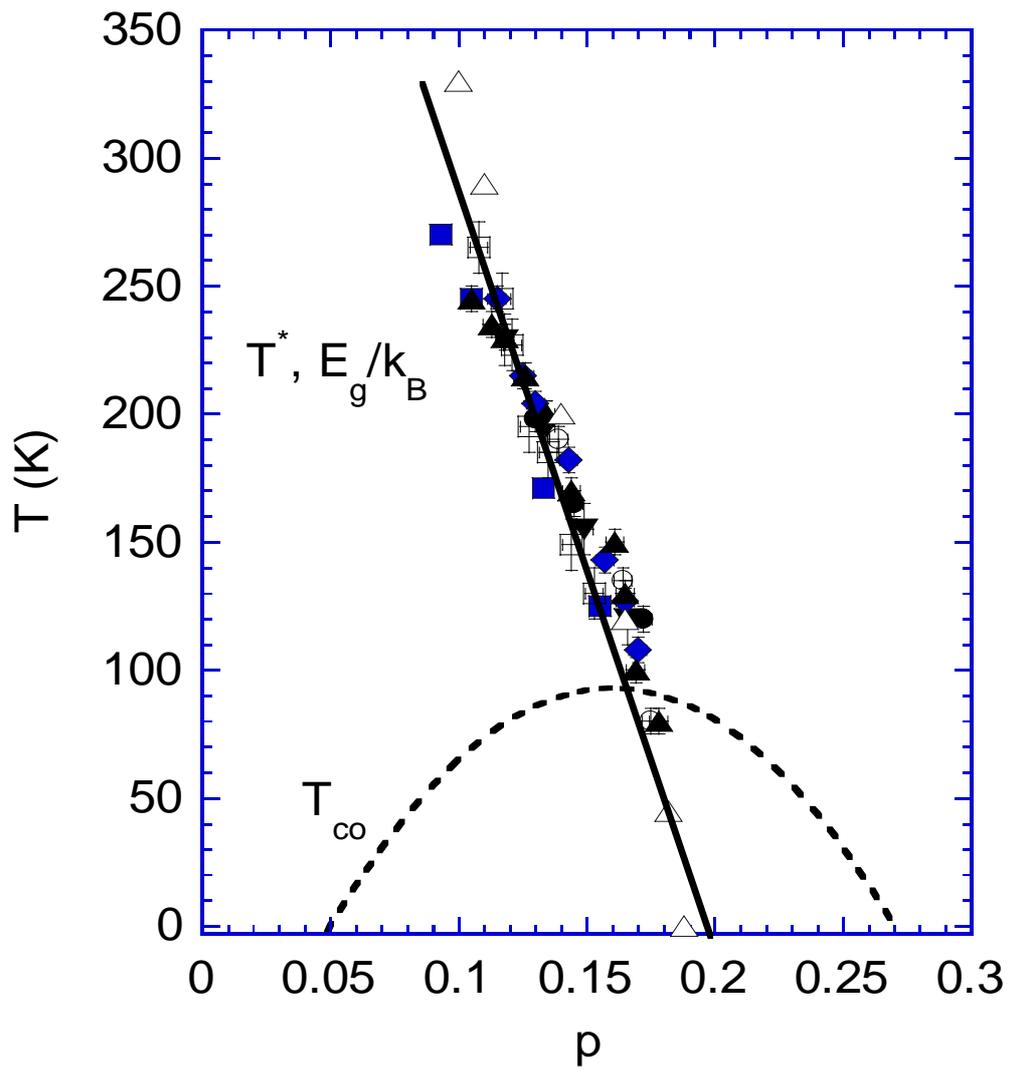